# Recommending Targeted Strangers from Whom to Solicit Information on Social Media

**Jalal Mahmud      Michelle X. Zhou      Nimrod Megiddo      Jeffrey Nichols      Clemens Drews**

IBM Research – Almaden

650 Harry Rd, San Jose, CA 95120

{jumahmud, mzhou, megiddo, jwnichols, cdrews}@us.ibm.com

## ABSTRACT

We present an intelligent, crowd-powered information collection system that automatically identifies and asks targeted strangers on Twitter for desired information (e.g., current wait time at a nightclub). Our work includes three parts. First, we identify a set of features that characterize one's willingness and readiness to respond based on their exhibited social behavior, including the content of their tweets and social interaction patterns. Second, we use the identified features to build a statistical model that predicts one's likelihood to respond to information solicitations. Third, we develop a recommendation algorithm that selects a set of targeted strangers using the probabilities computed by our statistical model with the goal to maximize the overall response rate. Our experiments, including several in the real world, demonstrate the effectiveness of our work.

**ACM Classification Keywords**

H.5.2. [Information Interfaces and Presentation]: User Interfaces

## INTRODUCTION

Hundreds of millions of social media posts (e.g., tweets) are generated daily. Not only does the buzz of the crowd provide a great deal of information [12, 28, 36], but also creates a unique opportunity for building a new type of crowd-powered information collection systems. Such systems will *actively* identify and engage the right people at the right time on social media to elicit desired information. Assume that one wants to know the current wait time at a local popular restaurant. The system would then send such a request to those who are *able* (i.e., people tweeted being near the restaurant), *willing*, and *ready* to provide the wait time information. These systems have several advantages over other crowd-powered systems, such as Social Q&A [17].

First, it can collect information from people beyond one's own social network, especially when no one in their network may possess the desired information (e.g., no friends happen to know the wait time at a specific restaurant). Second, it can collect timely and potentially more accurate information from a crowd at an opportune time (e.g., people at an event) instead of passively waiting for opted-in information [3, 4, 5]. Third, it can collect balanced information from a diverse population by explicitly distributing questions across people with different perspectives, such as liberals and conservatives.

Although the systems described above have great potential and a recent study also demonstrates the feasibility of manually soliciting information from strangers on social media [27], there is so far little research in this space. To capitalize on the potential value of this approach, we are building an intelligent, crowd-powered information collection system, called qCrowd. Given a user's information request, qCrowd aids the user in collecting the desired information in four steps. First, it monitors a social media stream (e.g., twitter stream) to identify relevant posts (e.g., tweets generated from foursquare.com indicating people at specific locations). Second, it evaluates the authors of identified posts and recommends a sub-set of people to engage (Figure 1a). Third, it generates questions and then sends them to each selected person (Figure 1b). Fourth, it analyzes received responses and synthesizes the answers together.

In this paper, our focus is on the second step, the process of *automatically* determining the targeted strangers to engage on social media for an information collection task. This process is non-trivial for several reasons. First, the system must determine a person's qualification for the task— whether a person has the *ability* to provide the requested information, such as being at a specific location or having knowledge of a particular product. This requires analyzing the content of one's social media posts, which is often a difficult natural language processing task. Second, the system must determine how *willing* a qualified person is to provide the requested information. We hypothesize that one's willingness to respond to strangers may be related to one's personality traits [30], such as friendliness. However, it is unknown what and how exactly various personal traits influence one's willingness. In addition, accurately predicting one's willingness to respond is critical to the success of qCrowd, since a previous study shows that unwilling people not only would not respond, but are also likely to mark the information request as a spam if they do not understand the value of the request, which may result in suspended operations [27]. Third, qCrowd must determine how *ready* a person is to respond at the time of request. As seen later, one's readiness may depend on a number of factors, including the temporal characteristics of one's behavior and their most recent activity on social media. Again, there is little investigation in computationally measuring what and how various factors impact one's readiness to respond.

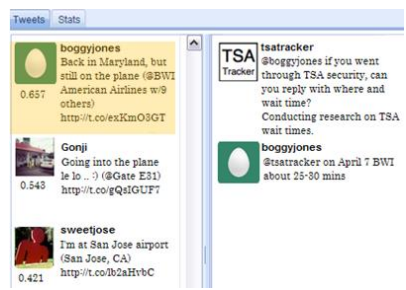

**Figure 1. A screenshot of qCrowd. (a) left panel: system-recommended three people, and (b) right panel: the question sent and response received.**

To address the challenges mentioned above, we build computational models to measure a person's *willingness*, *ability*, and *readiness* to respond based on one's behavior on social media. Since measuring one's ability is often domain or context dependent, currently we use a set of domain-specific rules to identify the targeted strangers who might possess the requested information. In the rest of the paper, we focus on the willingness and readiness models, which we believe are mostly domain agnostic.

Our work includes three parts. First, based on our real-world experiments and crowd-sourced surveys, we identify and compute a set of features that are likely to impact one's willingness and readiness to respond. These features capture various traits of a person, including one's personality and social interaction pattern. They are computed from several sources, including one's social media posts, social profile, and social media activities. Second, we train a statistical model to infer the contribution (weight) of each feature to one's willingness and readiness, which are then used to predict one's overall *likelihood to respond*. Third, we develop a recommendation algorithm that selects a subset of people based on the output of the statistical model. As explained later, our algorithm is designed to maximize the overall response rate of a request. Special care is thus needed to overcome the inherent imperfections (prediction errors) in any statistical models. To demonstrate the effectiveness of our approach, we have conducted extensive experiments including real-world, live experiments. Currently, we have implemented our models based on Twitter, which should be easily adapted to other social media platforms. As a result, our work offers three unique contributions.

- We build an extensible, feature-based model of predicting one's *likelihood to respond* to an information request on social media. Specifically, we identify a set of key features and their contributions (weights) to model a person's *willingness* and *readiness* to respond.

- We develop a recommendation algorithm that automatically selects a set of targeted strangers to maximize the overall response rate of an information request.

- We have applied qCrowd to real world scenarios. The results including demonstrated effectiveness provide validity and insights for building this new class of crowd-powered, intelligent information collection systems.

**RELATED WORK**

Our work is most closely related to a recent effort in studying the feasibility of directly asking targeted strangers on social media (Twitter) for information [27]. This work studies people's response patterns to questions in two domains and finds that it is feasible to obtain useful information from targeted strangers on social media with an average response rate of 42%. While their findings motivate our work, we advance their manual selection of targeted strangers by *automating* the selection process.

Our work is also related to a large number of efforts in online social Q&A services (e.g., [1, 2, 17]). Most of these services rely on volunteers to answer posted questions. To help get better answers faster, there are a number of efforts on automatically identifying suitable answerers for a given question [10, 20, 23, 24, 31, 40, 41]. Most of them however model the *ability* of potential answerers instead of their *willingness* or *readiness* to answer questions as we do.

Since we model a person's traits based on one's social media behavior, our work is related to a rich body of work on modeling people traits from social media, including location [13], demographics [33, 35], gender [35], personality [19], and political orientation [33, 35]. Like these works, we use one's social media posts and social networking activities (e.g., responding to other posts) to model personal traits, including personality and response behavior. Unlike these works, which model general traits of people, ours focuses on modeling the traits related to one's willingness and readiness to respond to a stranger's information requests on social media.

There is also work on studying and modeling people's responsiveness and availability in various situations (e.g., [8, 9, 26, 32, 38]). Closest to ours is studying response behavior in online social networks, such as Twitter [26, 32, 38]. While these studies focus on the behavior occurred *within* one's own social network [32, 38], ours examines how people would respond to *strangers* on an open social network. There is also work on modeling one's responsiveness to incoming instant messages [8] and availability [9] based on one's desktop computer activities. While we model similar traits of people (e.g., responsiveness and availability), our work differs in its use of one's social behavior, such as the content of their posts and social networking activities, which poses different technical challenges.

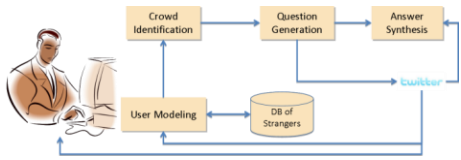

**Figure 2. qCrowd overview.**

## OUR SYSTEM
We provide an overview of qCrowd and its applications.

### Overview
As shown in Figure 2, qCrowd has four key components. The *user modeling* component creates a feature-based profile for each potential target on Twitter based on his/her tweeting activities, including the tweets posted. Each profile contains a set of computed features for modeling one's willingness, ability, and readiness to respond to a request. Such computed profiles are stored in a database. This component also updates the profile of a user who is already in the database. Based on a user's profile, the *crowd identification* component dynamically selects suitable targets to engage with. After a targeted person is identified, the *question generation* component generates a question suitable to the context (e.g., with a proper greeting) and then sends the question to the target via Twitter. The *answer synthesis* component parses and collates the received responses.

Unlike a traditional intelligent user interface that focuses on aiding human-computer interaction, qCrowd presents an intelligent user interface that facilitates human-computer-human interaction. In particular, it helps an information requester (e.g., a police officer) to better engage with a crowd (e.g., people happened to be near a crime scene) on social media to accomplish information solicitation tasks. The user interface of qCrowd is thus designed to facilitate such computer-mediated, human-human interaction. It allows a user to interact with qCrowd in one of three modes: (1) manual, (2) auto, and (3) mixed.

In the manual mode, a user (e.g., a marketer who wants to engage customers of a brand) watches a Twitter stream and manually selects individuals to send questions, composes the question, and synthesizes the answers received (Figure 3). In the auto mode, qCrowd performs these functions automatically but presents the results to the user (e.g., Figure 1). It can also run in a mixed mode, where one component (e.g., crowd selection) runs in the auto mode while another (e.g., question generation) runs manually. It always shows the user system-computed results (e.g., selected crowd) while allowing the user to interactively control the engagement process (e.g., editing and sending a question).

Our approach is easily applicable to different social platforms, as long as qCrowd has access to the needed information, such as one's social media posts and social interaction activities. It can then compute needed features and infer one's likelihood to respond to an information request. For example, in addition to our Twitter-based implementation, qCrowd technologies are also applied to enterprise social platforms, which support multiple types of social media activities, including online communities and forums.

### Current Applications
Since many people disclose location-based information (e.g., where they are) or their opinions/experiences about a product or event on social media, we have built two applications focusing on collecting two types of information on Twitter: location-based information (e.g., airport security check wait time) and people's opinions about a product or service (e.g., cameras, tablets, and food trucks). While each is general enough to represent a class of applications, the two are also distinct enough from each other, which allow us to test the generality of our approach.

Our work should also be easily extended to support other types of information collections. For example, we are applying qCrowd to information solicitations in crisis and emergency situations, since its true potential is to collect time-sensitive information from a diverse population. However, there may be new challenges that we have not encountered in our current applications. For example, we may need to consider privacy concerns if soliciting potentially sensitive information about people.

## DATA COLLECTION
To build a model for predicting one's likelihood to respond, we ran qCrowd in the manual mode to collect three data sets for both training and testing purposes. Our first two data sets were obtained in the process of collecting location-based information—airport security check wait time—via Twitter. Specifically, a human operator used qCrowd to manually identify people at an airport in the United States based on their tweets and then ask them for the airport security check wait time. Here is an example:

@bbx *If you went through security at JFK, can you reply with your wait time? Info will be used to help other travelers.*

Initially, one human operator asked 589 users and obtained 245 responses (42% response rate), which became our first data set *TSA-Tracker*-1. To collect more data in this domain, a different human operator then asked additional 409

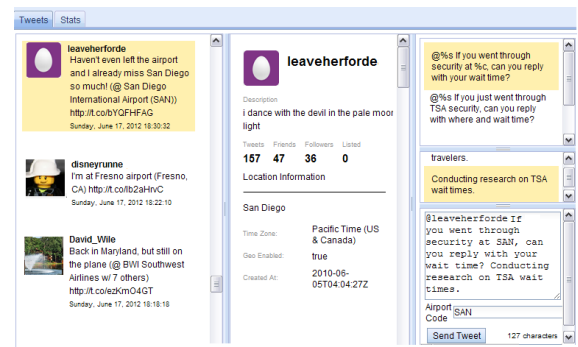

**Figure 3. qCrowd user interface in the manual mode. The left panel shows a set of filtered (by rules) tweets of those at an airport. The human operator examines the profile of a user and decides whom to engage (middle panel). The right panel allows the operator to compose questions to send.**

users and received 134 responses (33% response rate), which became our second data set, *TSA-Tracker*-2. Our third data set was obtained using the same method but by asking people on Twitter to describe their product/service experience. We asked people's opinions about three types of products: digital cameras, tablet computers, and food trucks. Here are two examples:

*@johnny Trying to learn about tablets...sounds like you have Galaxy Tab 10.1. How fast is it?*
*@bobby Interested in Endless Summer. Sounds like you have eaten there. Is the menu large or small (for a food truck)?*

A human operator asked 1540 users, and 474 responses were received (31% response rate). We refer to this data set *Product*. In each data set, we also collected the most recent tweets (up to 200 tweets) of each person engaged.

## BASELINES

The human operators managed to achieve an average response rate of 31-42% in a manual process. The process however was found quite difficult as it required the human operators to read the relevant tweets and identify the right people to engage in real time. We thus created two baselines to assess the difficulty of the process and potential value of our automated approach. Our first baseline is to engage random people on social media without any filtering. Our second baseline is to let end users (crowd) identify and select targeted strangers to engage on social media.

### Asking Random Strangers

To create a true baseline, we sent questions to random people on Twitter without considering their willingness, ability, or readiness to respond to these questions. Specifically, qCrowd *automatically* sent a question to a random person on Twitter at a fixed time interval (every 5 minutes). We created three different Twitter accounts and experimented with questions on three general topics: weather, public safety, and education. Here is an example question:

*@needy Doing a research about your local public safety. Would you be willing to answer a related question?*

Our aim was to send 250 questions on each topic. However, all three accounts were temporarily suspended by Twitter after sending a certain number of questions. On the weather topic, our system was able to send 187 questions and received only 7 responses. On public safety, 178 questions were sent and only 6 responses received. On education, only 3 responses were received after sending 101 questions. The response rates on all three topics were very low (well below 5% on each topic). This implies that it is ineffective to ask random strangers on social media without considering their willingness, ability, or readiness to answer questions. Moreover, the account suspension suggests that many people who received our questions may have flagged our account as a spamming account [27]. Thus, identifying the right people to engage is critical to the success of qCrowd.

### Crowd as Human Operator

Our second baseline was created by crowdsourcing a human operator's task to test a crowd's ability to identify the right targets to engage. We were also curious about the criteria that a crowd would use to identify targeted strangers.

To do so, we conducted two surveys on CrowdFlower, a crowd-sourcing platform [7]. The first survey asked each participant to predict if a displayed Twitter user would be *willing* to respond to a given question, assuming that the user has the ability to answer. The second survey asked each participant to predict *how soon* the person would respond assuming that s/he is willing to respond. The participant was also required to provide an explanation of their prediction. We deliberately designed two separate surveys to understand a participant's criteria for judging one's willingness and readiness, respectively and to avoid potential interferences.

*Willingness Survey*

The first survey had two settings depending on what information of a Twitter user was displayed to the participant. In the first setting, only the user's tweets were displayed. In the second setting, the user's twitter screen name, twitter profile URL, and tweets were all displayed. For each setting, we randomly picked 200 Twitter users, 100 each from the *TSA-tracker-1* and *Product* dataset. In each setting, we recruited 100 participants on CrowdFlower, each of which was given 2 randomly selected users for prediction. We measured the crowd's prediction accuracy by comparing their predictions with the ground truth—whether the users responded to our requests in our data collection process.

In the first setting (displaying tweets only), our participants were 29% correct in their prediction. In the second setting, they were 38% correct. Since everything was the same in both settings except the amount of information disclosed about a potential target, the additional information displayed in the second setting helped the prediction. These results suggest that it is also difficult for the crowd to identify targeted strangers to engage. By the comments of the participants who made correct predictions, several types of information were used in their prediction.

Most (56.7%) mentioned that they made their predictions by observing people' past interaction behavior: *"Talks to others and responds", and "The user seems extremely social, both asking questions and replying to others"*.

Some participants (10.45%) used one's activeness on Twitter as a predictor: *"This user tweets a lot, seems very chatty"*. Some participants (10.45%) used one's profile information: *"The user's job is Social Media representative..., his job involves answering questions"*; and *"being a social media guy and his tagline saying..."*

Several (6%) used one's re-tweeting behavior to predict: *"No (the person would not respond). Most of the tweets are retweets instead of anything personal"*. It is interesting to note that some were able to use one's personality (7.4%) for prediction: *"I think he won't respond. Doesn't seem to be very friendly"*.

| Responsiveness Features | Computation |
|---|---|
| Mean Response Time, Median Response Time, Mode Response Time, Max Response Time, Min Response Time | Avg(T), Med(T), Mod(T), Max(T), Min(T)<br><br>T denote previous response times computed from users' past responses to questions asked in Twitter |
| Past Response Rate | $N_R/N_D$, $N_R$ is the number of the user's previous responses and $N_D$ is the number of direct questions the user was asked in Twitter. |
| Proactiveness | $N_R/N_I$, $N_R$ is the number of user's responses and $N_I$ is the number of indirect questions the user was asked in Twitter, where an indirect question does not contain mentions of the user's name |

**Table 1. Responsiveness features and their computation.**

We also examined how often the above types of information were used by participants whose predictions were *incorrect*. We obtained similar results: users' past responsiveness (identified by 41% of such participants), one's activeness (7%), profile information (6%), re-tweeting behavior (4%), and personality (3%), respectively. Since the participants used a similar set of factors regardless their prediction results, this suggests that none of the factors alone is perfect for identifying a potential responder.

*Readiness Survey*

Our second survey asked each participant to judge how *soon* a person would respond to an information request, assuming that the person would respond. We used a multiple choice question with varied time windows as choices (e.g., within an hour and within a day). In this survey, we randomly chose 100 people from our collected data sets (50 from *TSA-Tracker-1* and 50 from *Product*). We recruited 50 participants on CrowdFlower, each of which was given two randomly chosen people and their twitter handlers. We computed prediction correctness by comparing participants' predictions with the ground truth. For example, if a participant predicted that person *X* will respond within an hour, however, the response was not received in time, the prediction is then incorrect. Our participants were 58% correct. Through their comments, the participants identified several factors that helped them in their predictions: 25% thought that one's activeness (frequent users) and steadiness (consistent usage patterns) on Twitter were good indicators; 30% of them considered the promptness of a person's response on Twitter a good predictor.

Overall, our crowd-sourcing experiments were very valuable. We learned that it is difficult for an end user to identify the right targets to engage, which justifies our effort to automate the process by recommending suitable targets. Moreover, our participants used several criteria, including one's personality and responsiveness, to identify the right targets. These criteria provided us with an empirical base (feature categories) to build our computational model.

## FEATURE EXTRACTION

To model one's willingness and readiness to respond to information solicitations on Twitter, we have identified five categories of features: *Responsiveness*, *Profile*, *Personality*, *Activity*, and *Readiness*. These features are mainly derived from one's tweet content, twitter profile, and social activities on Twitter. The first four categories, modeling a user's willingness to respond, were mainly identified by the participants in our first survey. The last category, modeling one's readiness to respond, was based on our second survey results but augmented to include additional features.

### Responsiveness

According to our survey participants, one's response behavior on Twitter was the top factor for predicting one's likelihood to respond. We have thus identified two key features to measure one's overall responsiveness based on his/her tweeting behavior. One is *response time*, measuring how *quickly* a person normally responds to a post directed to him/her on Twitter. Our hypothesis is that the faster the person responds, the more likely the person is willing to do so. The other feature is *proactiveness*, measuring how active usually a person is to respond to any posts on Twitter including the ones not directed to him/her. The more proactive a person is, the more likely the person is willing to answer a request. Table 1 shows the two features and their variants. To compute the value of these features, qCrowd collects each person's interaction history, including one's responses to other individuals on Twitter.

### Profile

Based on one's profile, our survey participants assessed the *socialness* of a person and used it as a predictor. We thus compute *CountSocialWords*, the frequency of social words appearing in a profile, including "*talking*" and "*communication*", adopted from the LIWC social process category [37]. We also added a set of words related to modern social activities, such as "*tweeting*" and "social network".

### Personality

Some of our survey participants mentioned personality being one of the factors for predicting one's likelihood to respond. We hypothesize that one's personality traits, such as friendliness and extraversion, may be related to one's willingness to respond to strangers. Since it is impractical to ask a stranger on social media to take a personality test, we need to derive one's personality from his/her social media behavior. Several researchers have found that word usage in one's writings, such as blogs and essays, can be used to infer one's personality [16, 18, 19, 21, 25, 37]. Inspired by the existing work, we used the LIWC dictionary to compute one's personality traits [34, 37]. LIWC defines 68 different categories, each of which contains several dozens to hundreds of words [34]. For each person, we compute a LIWC-based personality as follows.

| Readiness Features | Computation |
|---|---|
| Tweeting Likelihood of the Day | $T_D/N$, where $T_D$ is the number of tweets sent by the user on day D and N is the total number of tweets. |
| Tweeting Likelihood of the Hour | $T_H/N$, where $T_H$ is the number of tweets sent by the user on hour H and N is the total number of tweets. |
| Tweeting Steadiness | $1/\sigma$, where $\sigma$ is the standard deviation of the elapsed time between consecutive tweets of users, computed from users' most recent K tweets (where K is set, for example, to 20). |
| Tweeting Inactivity | $T_Q - T_L$, where $T_Q$ is the time the question was sent and $T_L$ is the time the user last tweeted. |

**Table 2. Readiness features and their computation.**

Let $g$ be a LIWC category, $N_g$ denotes the number of occurrences of words in that category in one's tweets and $N$ denotes the total number of words in his/her tweets. A score for category $g$ is then: $N_g/N$.

We exclude one's re-tweets when computing LIWC-based personality, since re-tweets are the content generated by others. Besides LIWC categories, psychologists have developed several personality models. One of the most well-studied is the Big5 personality model [15, 29]. It characterizes a person's traits from five aspects (also known as OCEAN): *openness*, *conscientiousness*, *extraversion*, *agreeableness*, and *neuroticism*.

Previous work [16, 18, 19, 21] has revealed correlations between Big5 personality [15] and LIWC-based features. More recently, Yarkoni [39] shows that correlations also exist between LIWC features and lower-level facets of Big5 [15]. Motivated by these findings and based on the word use in one's tweets, we compute one's additional personality features: Big5 traits and their lower-level facets.

To derive personality scores for each of the Big5 dimensions and their lower-level facets, we use the coefficients of correlation between them and LIWC categories [39]. We use a linear combination of LIWC categories (for which correlation was found statistically significant by Yarkoni [39]), where correlation coefficients are used as weights.

We extract a total of 103 personality features: 68 LIWC features, 5 Big5 traits, and 30 Big5 lower-level facets.

**Activity**
As indicated by our survey participants, one's demonstrated activity pattern on Twitter could signal one's willingness to respond. Intuitively, the more active a user is, the more likely the person will respond to a request. To capture this intuition, we compute two features: the total number of posts made overall (*MsgCount*) and daily (*DailyMsgCount*). These features were also used in prior works [14, 22] and help distinguish "sporadic" vs. "steady" activeness. We hypothesize that more "steady" users are more dependable and are more likely to respond when asked.

Several of our participants also noted re-tweeting as a factor to determine one's likelihood to respond. We thus incorporate one's retweeing behavior by two features: retweeting ratio (*RetweetRatio*), the ratio between the total number of retweets, and the total number of tweets sent, and daily retweeting ratio (*DailyRetweetRatio*), the ratio between the total number of retweets and the total number of tweets sent daily. These features help us distinguish users who create their own content vs. those who mostly retweet others' [11, 14]. We hypothesize that users who create their own content more often may be more willing to respond to a question, since responding involves content creation.

**Readiness**
Even if a person is willing to respond to a request, s/he may not be ready to respond at the time of request due to various reasons. For example, the person is in the middle of something and unable to respond immediately; or the person is experiencing a device malfunction (e.g., a dying battery) and unable to respond. The ability to predict one's readiness becomes more important when it requires time-sensitive responses, e.g., an information request in a crisis situation. However, one's readiness is highly context dependent and often difficult to capture as shown in our survey.

We thus use several features to approximate one's readiness based on the factors identified by our survey participants. Table 2 lists all the readiness features and their computation. Our rationale of choosing this set of features is twofold. First, these features are good indicators of one's readiness from a particular aspect. For example, *Tweeting Inactivity* implies one's unavailability. A larger value suggests either that the person is busy and hence uninterruptible, or that s/he is out of reach. Second, these features are easy and fast to compute as only one's tweeting activity is used.

**Feature Analysis**
In total, we identify 119 features to model one's willingness and readiness to respond to a request on social media. Since we do not expect all features to contribute to the models equally, we performed a series of Chi-square tests with Bonferroni correction to identify statistically significant features that distinguish people who responded from those who did not. Using our *TSA-Tracker*-1 data set, we found 45 significant features (False Discovery Rate was 2.8%). In *TSA-Tracker*-2, we found 13 significant features (False Discovery Rate was 11.2%). For the *Product* data set, 33 features were identified as significant (False Discovery Rate was 4.2%). Table 3 shows a common set of five significant features belonging to four feature categories across all three data sets. We have also run extensive experiments with various feature combinations to hope to find a combination that has significantly discriminative power. We discovered a combination of four top features: *communication* (a LIWC feature), *past response rate*, *tweeting inactivity*, and *tweeting likelihood of the day*. As shown later, this combination produces very impressive prediction results.

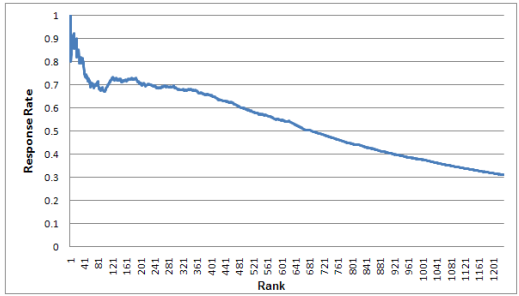

**Figure 4. Response rate in the training set with rank of individuals. A response rate R$_i$ is obtained when top i individuals are asked from the training set.**

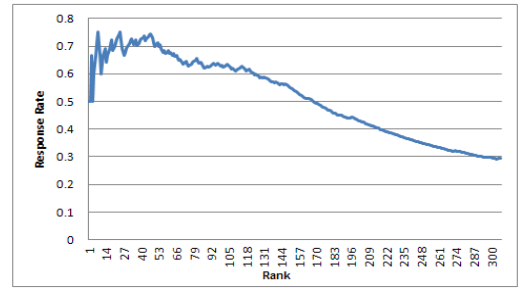

**Figure 5. Response rate in the test set with rank of individuals. A response rate R$_i$ is obtained when top i individuals are asked from the test set.**

## STATISTICAL MODELS

Using the features identified above, we build statistical models that predict a person's likelihood to respond. To do so, we partition each of our data sets randomly into *K* parts for *K*-fold cross validation. For each person in a training set, we compute all his/her features. The features and the person's response information (responded or not-responded) are then used to train a statistical model. Given a person in a test set and his/her computed features, the trained model outputs a probabilistic score that estimates the likelihood for the person to respond.

Formally, let $x_i$ be the feature vector of the *i*th person in a data set, and $y_i$ be the response label, such that if the person responded, $y_i = 1$; otherwise $y_i = 0$. In a simplified model, we assume that there is a unit benefit *B* of receiving a response and a unit cost *C* of sending a question. However, in reality, this assumption may not suffice. For example, the benefit of receiving additional answers to the same question may diminish as the number of answers grows. To build an accurate prediction model, we want to minimize prediction errors. Depending on the application, one type of error may be more costly than the other. For example, in one of our applications, the cost of sending a question may be much smaller than the benefit of receiving an answer. In this case, false negatives (e.g., missed people who would have responded) are more costly than false positives (e.g., selected people who do not respond).

However, most of the models typically do not differentiate two kinds of misclassification errors: false positive and false negative. By default, the classifier-building algorithm assumes that the designer wishes to minimize the overall misclassification rate. Similarly, a standard regression model does not distinguish between overestimating and underestimating probabilities, which in turn results in different types of misclassification errors.

Given the above considerations, we handle the difference in cost by properly weighing the training samples, assuming the unit cost and unit benefit. We weigh our training examples as follows. Positive examples (i.e., people who responded) are weighted by *B* − *C*, whereas negative ones (i.e., those who did not respond) are weighted by *C*. In other words, a false negative error, which is a misclassification of a positive example, incurs a missed-opportunity cost of *B* − *C*, where a false positive, which is a misclassification of a negative example, incurs a cost *C*.

| *Past Response Rate* (Responsiveness), *Tweet Inactivity* (Readiness), *Excitement-Seeking* (Personality), *Cautiousness* (Personality), *DailyMsgCount* (Activity) |
|---|

**Table 3. Statistically significant features and their category common to all three data sets.**

We have used two popular statistical models in WEKA [6]: Support Vector Machines (SVM) and Logistic Regression, to predict the probability for a person to respond.

## RECOMMENDATION ALGORITHM

Our trained statistical model can be used in various ways to recommend a set of people to engage. The simplest is to use it as a binary classifier, which just predicts a person as a *responder* or *non-responder*. qCrowd can then engage those who are classified as responders. Alternatively, qCrowd can rank the people by the probability scores produced by the statistical models and then select the top-*K* to engage. However, due to inherent imperfections (prediction errors) in any prediction models, in reality, the predicted top-*K* people may not necessarily be the best choices who will maximize the overall response rate. To maximize the overall response rate for a given set of candidates, we thus have developed the following algorithm.

For each candidate *i*, our models compute a probability $p_i$. We then sort all candidates by the corresponding probabilities in a decreasing order: {$p_1,…,p_n$}. We focus on intervals in this linear order rather than looking at all possible subsets of candidates. The justification for this restriction is that the linear orders generated by the models exhibit a good correlation with response rates. Figure 4 shows that in general the higher rank of an individual is, the higher the average response rate is (response rate achieved by asking all individuals with a higher rank). Here one's rank is determined by his/her computed probability. A person ranked first means his likelihood to respond is greater than or equal to that of rest. Figure 5 shows a similar curve when the model generated from a training set is applied to a test set. While both show that the probability to respond helps predict response rates, they also show that the top segment containing the highest ranked candidates may *not* produce the overall maximal response rate.

|           | TSA-tracker-1 | | TSA-tracker-2 | | Product | |
|-----------|:---:|:---:|:---:|:---:|:---:|:---:|
|           | SVM | Logistic | SVM | Logistic | SVM | Logistic |
| **Precision** | 0.62 | 0.60 | 0.52 | 0.51 | 0.67 | 0.654 |
| **Recall** | 0.63 | 0.61 | 0.53 | 0.55 | 0.71 | 0.62 |
| **F1** | 0.625 | 0.606 | 0.525 | 0.53 | 0.689 | 0.625 |
| **AUC** | 0.657 | 0.599 | 0.592 | 0.514 | 0.716 | 0.55 |

**Table 4. Performance of prediction models.**

Thus, our approach involves selecting an interval [$i$, $j$] ($1 \leq i < j \leq n$) from the training set, where the corresponding interval subset $\{p_i,…,p_j\}$ produces an overall maximal response rate among all interval subsets. In our selection, we ignore short intervals at the top of the ranking. The rationale is that the variance in these small sets is large. A short interval that appears to produce a high response rate in the training set does not imply that the corresponding interval in the test set also has a high response rate. We also experimented with a restricted choice of intervals: choosing only those that extend to the top [$i$, $n$]. We have observed that restricting to intervals [$i$, $n$] rather than [$i$, $j$] produces sub-optimal results. The best subinterval [$i_r$, $j_r$] in the training defines a corresponding sub-interval [$i_s$, $j_s$] in the test set, based on percentiles. That is, if $m$ is the cardinality of the test set, then $i_s = [(i_r \cdot m)/n]$ and $j_s = [(j_r \cdot m)/n]$.

We can also incorporate additional constraints in our interval selection. For example, if a minimum size of the interval is specified, our method will ignore intervals that are smaller than the specified minimum.

## EVALUATION

To test the performance of our predictive models and recommendation algorithm, we have conducted an extensive set of experiments.

### Evaluating Prediction Model

One key goal is to predict the *likelihood* for a stranger on social media to respond to our information requests. To evaluate the performance of our prediction models, we adopted a set of standard performance metrics including precision, recall, F1, and AUC (Area under ROC curve). For each data set, we performed 5-fold cross validation experiments with uniform weights, where $B - C = C = 1$.

Table 4 shows the results for different data sets using SVM and Logistic Regression. Overall, the SVM-based model outperformed logistic regression. The SVM-based model for the *Product* data set also achieved the highest AUC value (0.716). Since the SVM-based model performed better in our experiments, we report various results obtained using this model in the rest of the paper.

|   | TSA-tracker-1 | TSA-tracker-2 | Product |
|---|:---:|:---:|:---:|
| Baseline | 42% | 33% | 31% |
| Binary classification | 62% | 52% | 67% |
| Top-K selection | 61% | 54% | 67% |
| Our algorithm | 67% | 56% | 69% |

**Table 5. Comparison of average response rates for different recommendation approaches.**

| Interval Size | Optimal Interval | Response Rate | Recommendation Recall |
|:---:|:---:|:---:|:---:|
| 25% | [67%, 92%] | 76% | 37% |
| 50% | [46%, 96%] | 68% | 64% |
| 75% | [19%, 94%] | 53% | 82% |
| 100% | [0%, 100%] | 31% | 100% |

**Table 6. Response rate and recall by our recommendation algorithm with fixed size interval (Product Data, all features, SVM Model).**

### Evaluating Recommendation Algorithm

As described earlier, our algorithm aims at selecting a targeted population to maximize the overall response rate. We thus compared the overall response rate achieved by qCrowd against our baseline. Our baseline is the response rates achieved by a human operator during our data collection process. For testing purpose, we used "*asking at least K% of people from the original set*" as a constraint to search for the interval that maximizes the response rate. We computed the response rates on all three data sets with varied $K$ (e.g., $K$=5%, …, 90%) to find the respective optimal intervals. For comparison purpose, we also computed the response rates achieved using a simple binary classification (response rate is the precision of the predictive model) and simply selecting the top-$K$ (e.g., $K$=5%, …, 90%.) people by their computed probabilities.

Table 5 shows the results achieved by different approaches. Our algorithm outperformed others in all three data sets.

#### Evaluating Recall of Recommendation

Although our goal is to maximize the overall response rate, we are also interested in finding out what percentage of actual responders that qCrowd can identify. This essentially measures the "recall" of our algorithm. When using a top-$K$ selection, on average, the recall was 57%, 45%, and 63% for the three data sets, respectively. Using our recommendation algorithm, on average over varied $K$ (the minimal percentage of people to ask), the recall was 57%, 46%, and 65%. Since the recall is affected by the number of people are actually engaged, we performed a set of experiments with varied, fixed interval sizes. As shown in Table 6, as the fixed interval size (in percentiles) increases, the response rate decreases and the recall improves. In practice, we want to find an "optimal" interval size that achieves a balanced high response rate (among the recommended most people would respond) and high recall (most of those who are likely to respond were recommended).

#### Evaluating the Use of Different Feature Sets

We also wanted to see the effect of different feature uses on response rates. Table 7 shows various response rates using our algorithm with different feature sets when the minimum percentage to ask was set at 5%. The set of statistically significant features performed the best. It is notable that the top-4 features that we discovered (*communication*, *past response rate*, *tweeting inactivity,* and *tweeting likelihood of the day*) produced the best response rate for the *Product* data set and also performed well across all the data sets. Commonly significant features across all 3 data sets (listed in Table 3) also performed reasonably well. This implies that our features are fairly domain agnostic.

| Feature Set | Response Rate | | |
|---|---|---|---|
| | TSA-tracker-1 | TSA-tracker-2 | Product |
| All | 0.79 | 0.72 | 0.78 |
| Significant | 0.83 | 0.75 | 0.82 |
| Top-10 Significant | 0.83 | 0.74 | 0.81 |
| Top-4 features | 0.82 | 0.73 | 0.83 |
| Common significant features | 0.81 | 0.72 | 0.82 |

**Table 7. Response rates using our algorithm for different feature uses, minimum interval was 5%, SVM Model.**

When we tested on varied *K* values (e.g., *K*=5%, …, 90%. etc.), we found that use of statistically significant features yields average response rates of 69%, 66%, and 72% for *TSA-tracker-1*, *TSA-tracker-2*, and *Product* data sets, respectively. These results show a significant improvement over our two baselines: (1) the response rate (<< 5%) achieved by randomly sending questions to strangers, and (2) the response rate (31-42%) achieved by a human operator during the data collection process.

*Domain Sensitivity*
We randomly selected 500 users from *TSA-tracker-1*, built a SVM-based model using commonly significant features, and then applied it to the *Product* data set and vice versa. The average response rates (with minimum interval size 5%) were 70% for *TSA-Tracker-1* and 64% for *Product.* This shows that our models are fairly domain independent.

*Varied Cost and Benefit*
So far we presented results with statistical models trained with equally weighed positive (responded) and negative (not-responded) examples, where B – C = C = 1. To test how our models perform with varied benefit and cost, we repeated our experiments with different benefit (B) to cost (C) ratio with C = 1. We observed varied response rates using our SVM-based model (Table 8). We noticed the slight improvement of response rates with the increased benefit (B) to cost (C) ratio. This validates our hypothesis that in a system like ours minimizing misclassifications of false negative errors improves prediction.

**Live Experiments**
In addition to evaluating the effectiveness of our work on previously collected data sets, we also conducted live experiments. qCrowd was set on the auto mode and dynamically selected strangers on Twitter to send them questions to collect security check wait time at different airports and product experience. In the first case, qCrowd used Twitter's Search API and a set of rules to first identify 500 users who tweeted about being at any US airport. From this set, it randomly asked 100 users for the security wait time. From the remaining 400 users, qCrowd used the recommendation algorithm to identify 100 users for questioning. It used the SVM-based model with the identified significant features of *TSA-tracker-1*. We waited 48 hours for the responses. 29 of the 100 users from the first set responded (29% response rate) and 66 out of the 100 users from the second set responded (66% response rate). The same process was repeated for sending product-related questions. The only difference was that the initial set of 500 users was identified using Twitter's Search API and a set of rules to detect users who had tweeted about digital cameras, tablet computers, or food-trucks. Response rate by random questioning was 26% and our algorithm achieved 60%. Thus, our live experiments also demonstrated the effectiveness of our work in a real world setting.

**DISCUSSIONS**
Compared to traditional social Q&A systems, our work offers two unique advantages: *activeness* and *timeliness* in collecting desired information. qCrowd *actively* seeks out and solicits information from the right people at the *right time* on social media. It is particularly suitable for collecting *accurate* and *up-to-date* information about specific situations (e.g., crowd movement at an airport) from the people who are or were just in the situation. However, the dynamic and complex nature of social media also poses unique challenges for us. Below we discuss several limitations of our current work and corresponding design implications for building this new class of systems.

**Skew in the User Base**
Our current model favors active users on social media. These users are associated with a higher level of willingness to be engaged and hence are more likely to be targeted by our system. As a result, the selected targets may be skewed. On the other hand, it is difficult to create a model that "promotes" inactive users due to a lack of information on these users. If not careful, engaging inactive, unknown users could jeopardize the success of the whole operation as demonstrated by our experiments. Therefore, our ongoing work is to develop methods that can identify "inactive" users, e.g., those similar to active ones in comparable aspects like personality instead of their activity patterns.

| Benefit to Cost Ratio | TSA-Tracker-1 | TSA-Tracker-2 | Product |
|---|---|---|---|
| 2 | 0.83 | 0.75 | 0.82 |
| 5 | 0.85 | 0.79 | 0.84 |
| 10 | 0.87 | 0.81 | 0.87 |
| 100 | 0.88 | 0.82 | 0.87 |

**Table 8. Response rates achieved w/ different benefit to cost ratio, minimum interval 5%, SVM with significant features.**

**Modeling the Fitness of a Stranger to Engage**
Currently, we focus on modeling one's likelihood to respond, based on this person's ability, willingness, and readiness to respond to a request on social media. However, our current model does not consider one's traits that may impact the overall *quality* of received responses or cause unexpected viral effects. For example, a more dutiful and trustworthy person may provide a higher quality response, while one's hatefulness might prompt malicious behavior on social media. Moreover, information exchange in qCrowd occurs mainly between the asker (the system) and answerer (a stranger), without any moderation by a larger group. This removes the potential reputation and filtering benefits of typical Social Q&A sites, like Quora, to govern the quality of crowd-sourced information. To ensure the quality of responses and prevent potential mischief, we thus must assess the overall *fitness* of a stranger to engage on social media. Beyond modeling one's likelihood to respond, a fitness model should include people traits that impact various aspects of an information collection process on social media, such as the quality of responses received and potential influence of various response behaviors.

**Handling Complex Situations**
Our current goal is to recommend the right set of strangers to achieve a particular response rate. While this can handle many real-world applications, it does not cover complex information collection situations. Suppose that our goal is to collect at least *K* data points (e.g., wait times) each hour at an airport. In practice, for a given hour, there may be not enough candidates—people who were at the airport and tweeted about their location. To handle such situations, we must model and estimate the size of the overall candidate pool in various situations. This may involve estimating the probability of people appearing at the desired location or the probability of publicizing their location on social media. In addition, different applications may have different cost and benefit measures of sending a request and getting a response, respectively. Such benefit or cost may also change as the responses are received. Assume that the goal is to obtain opinions from multiple people. The benefit of getting additional answers diminishes as the received answers start to converge. To systematically handle various constraints, an optimization-based model is then needed to incorporate all the constraints (including benefit and cost) into one objective function, which computes the net benefit. The goal of our recommendation is then to find a targeted population such that the net benefit is maximized.

**Handling Unexpected Answers**
Our goal is to ask a targeted population on social media to respond to our requests. Due to the public nature of social media, people who are *not* asked can still see our posted questions and may volunteer to provide us with an answer. For example, when collecting our *Product* data set, an owner of a food truck unexpectedly responded to one of our questions. Our current approach has not taken such situations into account. It would certainly be interesting to incorporate such situations into our methodology. This may also become a way to "grow" potential targets.

**Protecting Privacy**
Although people volunteer information about themselves on social media, they may not be fully aware how the disclosed information can be used or have complete control over the shared information. We have not conducted in-depth studies on how the people who were selected feel about the process and what their concerns might be. With such information, we can better tune our selection algorithms to exclude people who wish not to be engaged.

## CONCLUSIONS
As hundreds of millions of people reveal themselves daily on social media like Twitter, our current research is on how to engage the right people at the right time on social media to crowd-source desired information. To achieve this goal, we are building qCrowd, an intelligent, crowd-powered information collection system. In this paper, we focus on qCrowd's ability to automatically identify and select the right strangers to engage on Twitter. Specifically, we have presented how we model and compute a stranger's likelihood to respond to an information request on social media.

Our current model measures one's willingness and readiness to respond based on a set of features derived from his/her social behavior on Twitter. Using this model, not only can we predict one's probability to respond to a question, but we have also identified a sub-set of features that have significant prediction power. Moreover, we have presented a recommendation algorithm that can automatically select a set of targeted strangers to engage based on their likelihood to respond with the goal of maximizing the overall response rate of the selected population. To validate our work, we have conducted an extensive set of experiments including live ones in a real-world setting. Our experiments demonstrate the effectiveness of both our prediction model and recommendation algorithm. In particular, our results show that qCrowd can achieve an average response rate over 65% in our two application domains, a significant improvement over the response rates achieved by a human operator in the range of 31-42%.

## ACKNOWLEDGEMENTS
Research was sponsored by the U.S. Defense Advanced Research Projects Agency (DARPA) under the Social Media in Strategic Communication (SMISC) program, Agreement Number W911NF-12-C-0028. The views and conclusions contained in this document are those of the author(s) and should not be interpreted as representing the official policies, either expressed or implied, of the U.S. Defense Advanced Research Projects Agency or the U.S. Government. The U.S. Government is authorized to reproduce and distribute reprints for Government purposes notwithstanding any copyright notation hereon.

## REFERENCES
1. WarriorForum – Internet Marketing Forums, http://www.warriorforum.com/

2. Slickdeals, http://slickdeals.net/forums/forumdisplay.php?f=19
3. Quora, http://www.quora.com/
4. Yahoo Answers, http://www.answers.yahoo.com
5. Facebook Questions, http://www.facebook.com
6. WEKA. http://www.cs.waikato.ac.nz/ml/weka/
7. Crowdflower. http://crowdflower.com/
8. Avrahami, D. and Hudson, Scott E. 2006. Responsiveness in Instant Messaging: Predictive Models Supporting Inter- Personal Communication, In *Proc. CHI 2006*.
9. Begole, J., Tang, J.C., Smith, R.E., and Yankelovich, N. 2002. Work rhythms: Analyzing visualizations of awareness histories of distributed groups. In *Proc. CSCW'02*.
10. Bouguessa, M. Dumoulin, B., and Wang, S. 2008. Identifying authoritative actions in question-answering forums: the case of Yahoo! answers: In *Proc. KDD 2008*.
11. Boyd, D. Golder, S., Lotan, G. 2010. Tweet, Tweet, Retweet: Conversational Aspects of Retweeting on Twitter. In *Proc.HICSS 2010*.
12. Chakrabarti, D., Punera, K.2011., Event Summarization using Tweets, In *Proc. ICWSM 2011*.
13. Cheng, Z., Caverlee, J., and Lee, K.2010. You Are Where You Tweet: A Content-Based Approach to Geo-locating Twitter Users. In *Proc. CIKM 2010.*
14. Castillo, C., Mendoza, M., and Poblete, B. 2011, Information Credibility on Twitter. In *Proc. WWW 2011*.
15. Costa, P.T.,Jr. & McCrae, R.R. 1992. Revised NEO Personality Inventory (NEO-PI-R) and NEO Five-Factor Inventory (NEO-FFI) manual. Odessa, FL: Psychological Assessment Resources.
16. Fast, Lisa A.; Funder, David C. 2008., Personality as manifest in word use: Correlations with self-report, acquaintance report, and behavior. Journal of Personality and Social Psychology, *Vol 94(2), 2008*.
17. Gazan, R (2011). Social Q&A. J. of the Amer. Soc. for Info Science & Technology, 62(12), 2301-2312.
18. Gill, A.J., Nowson, S., Oberlander, J. 2009. What Are They Blogging About? Personality, Topic and Motivation in Blogs, In *Proc. ICWSM 2009.*
19. Golbeck, J. Robles, C., Edmondson, M., Turner, K. 2011. Predicting Personality from Twitter. In *Proc. IEEE SocialCom.*
20. Guo, J., Xu, S., Bao, S., and Yu, Y. 2008. Tapping on the potential of Q&A community by recommending answer providers. In *Proc. CIKM 2008*.
21. Hirsh, J.B. and Peterson, J.B., 2009, Personality and language use in self-narratives Journal of Research in Personality.
22. Lee, K., Eoff, B.D. and Caverlee, J.2011. Seven months with the devils: A long term study of content polluters on Twitter. In *Proc. ICWSM 2011*.
23. Liu, X., Croft, W. B., and Koll, M.2005. Finding experts in community-based question-answering services, In *Proc. CIKM'05*, 315-316.
24. Jurczyk, P., and Agichtein, E., 2007, Discovering authorities in question-answering communities by using link analysis. In *Proc.CIKM 2007*.
25. Mairesse, F., Walker, M. 2006. Words Mark the Nerds: Computational Models of Personality Recognition through Language, In *Proc. of CogSci 2006.*
26. Morris, M., Teevan, J., and Panovich, K. 2010. What Do People Ask Their Social Networks, and Why? A Survey Study of Status Message Q&A Behavior. In *Proc. CHI,2010.*
27. Nichols, Jeffrey, Kang, Jeon-Hyung, 2012, Asking Questions of Targeted Strangers on Social Networks, In *Proc. CSCW2012.*
28. Nichols, J., Mahmud, J., Drews, C., 2012, Summarizing Sporting Events from Twitter., In *Proc. IUI2012.*
29. Norman,W. T. 1963. Toward an adequate taxonomy of personality attributes: replicated factor structure in peer nomination personality rating. Journal of Abnormal and social Psychology, 66:574-583.
30. O'Brien, T. and DeLongis, A. The interactional context of problem-, emotion-, and relationship-focused coping: The role of the big five personality factors. Journal of personality, 64(4):775–813, 1996.
31. Pal, A., Farzan, R., Konstan, Joseph A. and Kraut, Robert E,2011, Early Detection of Potential Experts in Question Answering Communities, In *Proc. UMAP 2011*.
32. Paul, S.A., Hong, L., and Chi, E.H., 2011. "Is Twitter a Good Place for Asking Questions? A Characterization Study", In *Proc. ICWSM'11 Posters.*
33. Pennacchiotti, M., Popescu, A.M. 2011. A Machine Learning Approach to Twitter User Classification, In *Proc. ICWSM 2011*.
34. Pennebaker, J.W., Francis, M.E., Booth, R.J., 2001, Linguistic Inquiry and Word Count. Erlbaum Publishers, Mahwah, NJ.
35. Rao, D., Yarowsky, D., Shreevats, A., Gupta, M. 2010, Classifying Latent User Attributes in Twitter, In *Proc. of SMUC'10*.
36. Sakaki, T., Okazaki, M., Matsuo, Y. 2010, Earthquake shakes twitter users: real-time event detection by social sensors, In *Proc. of WWW 2010.*
37. Tausczik, Yla R. and Pennebaker, James W, 2010. The Psychological Meaning of Words: LIWC and Computerized Text Analysis Methods, In. Journal of Language and Social Psychology 29(1) 24– 54
38. Teevan, J., Morris, M., and Panovich, K. 2011. Factors Affecting Response Quantity, Quality, and Speed for Questions Asked via Social Network Status Messages. In *Proc. ICWSM 2011*.
39. Yarkoni, Tal. 2010. Personality in 100,000 words: A largescale analysis of personality and word usage among bloggers. Journal of Research in Personality.
40. Zhang, J., Ackerman, M. S., and Adamic, L. 2007. Expertise networks in online communities: structure and algorithms. In *Proc. WWW, 2007*.
41. Zhou, Y., Cong, Z., Cui, B., Jensen, C.S., and Yao, J. Routing Questions to the Right Users in Online Communities. In *Proc. ICDE 2009*.